\documentclass[%
superscriptaddress,
preprint,
showpacs,
preprintnumbers,
amsmath,amssymb, aps, prx, 
floatfix,
12]{revtex4-1}
\usepackage{color}
\usepackage{graphicx}
\usepackage{dcolumn}

\usepackage{bm}
\usepackage{subfig} \usepackage{lineno} \usepackage{hyperref}
\usepackage{float}
\begin{document}
\preprint{APS/123-QED}
\title{Proposal of X-ray absorption spectroscopy and magnetic circular dichroism using broadband free-electron lasers}

\author{Bangjie Deng} 
\affiliation{School of Nuclear Science and Technology, Xi'an Jiaotong University, Xi'an, Shaanxi 710049, China}

\author{Jiawei Yan} 
\affiliation{Shanghai Institute of Applied Physics, Chinese Academy of Sciences, Shanghai 201800, China}
\affiliation{University of Chinese Academy of Sciences, Beijing 100049, China.}

\author{Qingmin Zhang}
\email{zhangqingmin@mail.xjtu.edu.cn}
\affiliation{School of Nuclear Science and Technology, Xi'an Jiaotong University, Xi'an, Shaanxi 710049, China}

\author{Yaodong Sang} 
\affiliation{School of Nuclear Science and Technology, Xi'an Jiaotong University, Xi'an, Shaanxi 710049, China}

\author{Haixiao Deng}
\email{denghaixiao@sinap.ac.cn} 
\affiliation{Shanghai Institute of Applied Physics, Chinese Academy of Sciences, Shanghai 201800, China}
\date{\today}

\begin{abstract}

X-ray free-electron lasers (XFELs) have been widely used for applications such as X-ray crystallography and magnetic spin probes because of their unprecedented performance. Recently, time-resolved X-ray magnetic circular dichroism (XMCD) with ultrafast XFEL pulses made it possible to achieve an instantaneous view of atomic de-excitation. However, owing to the narrow bandwidth and coherence of XFEL, X-ray absorption spectroscopy (XAS) and XMCD are time- and effort-consuming for both machine scientists and users of XFELs. In this work, an efficient scheme using a broadband XFEL pulse and single-shot X-ray spectrometer is proposed, in which the XAS and XMCD measurements can be accomplished with the same machine condition. An evolutionary multiobjective optimization algorithm is used to maximize the XFEL bandwidth offered by the Shanghai soft X-ray FEL user facility without additional hardware. A numerical example using MnO is demonstrated, showing that using approximately 1000 consecutive XFEL shots with a central photon energy of 650 eV and full bandwidth of 4.4\%, precise spectral measurements for XAS and XMCD can be achieved. Additional considerations related to single-shot XAS and XMCD are discussed.

\end{abstract}

.

\maketitle


\section{Introduction}

X-ray absorption spectroscopy (XAS) and X-ray magnetic circular dichroism (XMCD) with coherent light sources are two popular techniques used at the frontiers of materials science because they can probe element-specific electric and magnetic properties \cite{Dr1992NEXAFS,Stohr1999,Singhal2010}. Currently, synchrotron radiation is the leading available X-ray source for XAS and XMCD experiments; however, limited photons and inadequate temporal resolution of individual X-ray pulses do not support ultrafast dynamics and spintronics \cite{Radu2011}. Although high-harmonic generation from rare gas is suitable for time-resolved experiments, significant challenges remain in enhancing the pulse energy and stability of such X-ray sources \cite{Popmintchev2012,Fleischer2014}.

Owing to ultrafast pulse duration, well-defined polarization, and high brightness \cite{Emma2010,Allaria2014}, X-ray free-electron lasers (XFEL) are the most appealing light source for time-resolved XAS/XMCD experiments that capture instantaneous microscopic activity. Usually, the relative bandwidth of an XFEL is approximately 0.1\%, even in self-amplified spontaneous emission (SASE) modes \cite{Bonifacio1984}, which is not large enough to cover the entire XAS/XMCD spectra in one pulse. In the first time-resolved XAS/XMCD experiment using an XFEL \cite{Higley2016}, a narrow-bandwidth XFEL was used to scan the entire XAS/XMCD spectra within an energy band of 40 eV, which contrasts with the 0.1--0.2 eV typically covered with one XFEL shot after a grating monochromator. Time- and effort-consuming machine controls are necessary, such as the undulator gap tuning, beam orbit maintenance, and monochromator adjustment. Compared to synchrotron radiation light sources, limited experimental stations can be constructed for an XFEL facility. Therefore, such conventional methods cannot be widely used in the XAS/XMCD experiments due to their low operational efficiency.

In addition to pursuing narrow-bandwidth FEL pulses, the large-bandwidth operation mode of XFEL has been proposed in recent years \cite{song2018bandwidth, 19, 20,21,Prat2016,24}. Because one broadband XFEL pulse covers the entire energy range of absorption spectra and the number of X-ray photons in each pulse reaches $ 10^{12} $, in principle, the XAS and XMCD experiments can be accomplished without major interactions with XFEL machine components. In this study, an efficient XAS/XMCD scheme driven by broadband XFEL pulses is proposed. In order to determine satisfactory radiation bandwidths and corresponding operation parameters in the Shanghai soft-X-ray free-electron laser (SXFEL) user facility, an evolutionary multiobjective optimization algorithm is utilized to maximize FEL bandwidth. A numerical example is defined to demonstrate the XAS and XMCD method driven by broadband XFEL. In conventional XAS and XMCD experiments with monochromatic synchrotron radiation, absorption cross-sections are usually derived by measuring the fluorescence yield \cite{Goedkoop1997Soft,Lobenstine1980Further}, the current of Auger electrons \cite{Matsui2010}, and the $ E_{ph} $ spectrum of the transmitted X-ray\cite{Kotani2009,Miedema2013The,Higley2016}. However, the two former methods are not suitable for spectroscopy experiments with broadband FELs because their measured yields are only related to the flux of the XFEL pulses \cite{Ge2013Direct,Mathieu1985Auger}. In addition, the flight time of escaped Auger electrons can be modified by a strong magnetic field during XMCD measurement. In contrast, a method based on a transmission spectrum is practical for the XAS and XMCD experiments with broadband XFEL pulses because the flux and energy spectrum of the broadband XFEL can be obtained. Recently, several such methods have been implemented for absorption spectrum measurements \cite{Bernstein2009,Higley2016}, including the use of a grating raster for the simultaneous measurement of incident and transmitted $ E_{ph} $ spectra \cite{Bernstein2009}. However, in order to monitor the polarization properties of incident or transmitted FEL pulses, in this study, an e-TOF-based instrument was chosen because of its ability to simultaneously measure both the XFEL spectrum and polarization \cite{Zhang:2017ncj}. In our case, using an SXFEL operated at a 50-Hz repetition rate \cite{Ming-HaoSong:1709007} and an e-TOF-based high-resolution spectrometer \cite{Zhang:2017ncj}, the absorption spectrum of an MnO sample was measured in 1 min after careful calibration. Therefore, the broadband XFEL offers a more efficient method for both steady-state and time-resolved XAS and XMCD experiments.

This paper is organized as follows. In Section \ref{sec2}, the expectation of broad-bandwidth FEL pulses generation in the SXFEL is studied with an evolutionary multiobjective optimization algorithm, followed by a description of how a broadband XFEL can drive XAS and XMCD, and the details of the proposed experimental procedures in Section \ref{sec:proceduces}. In Section \ref{sec4}, using MnO as an example, numerical results and a corresponding sensitivity analysis are presented. Further discussion of single-shot XAS and XMCD are also discussed in Section \ref{sec5}. Finally, conclusions are summarized in Section \ref{sec6}.

\section{Broadband FEL pulse generation in SXFEL}
\label{sec2} 
\begin{figure*}[htp] 
\centering 
\includegraphics[width=\linewidth]{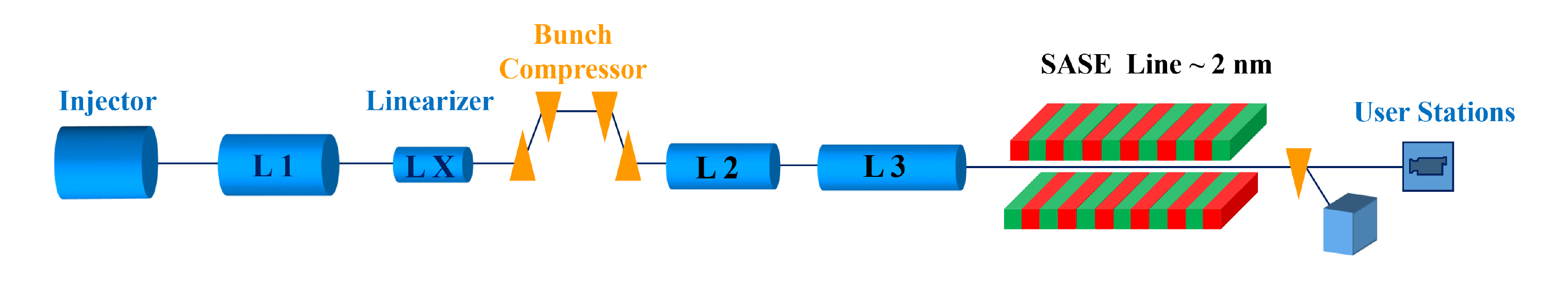}
\caption{Layout of the SXFEL user facility linac and the SASE line}
\label{FIG1} 
\end{figure*}

An evolutionary multiobjective optimization algorithm is used to find the maximum available radiation bandwidth in SXFEL, which is critical for XAS and XMCD experiments. According to the FEL resonance condition \cite{17}: 
\begin{equation} 
\lambda  = \frac{{{\lambda _u}}}{{2{\gamma ^2}}}(1 + \frac{{{K^2}}}{2}), 
\end{equation} 
where $\lambda $ is the radiation wavelength, ${\lambda _u}$ is the undulator period length, $\gamma $ is the average Lorentz factor of the electrons, and K is the undulator field parameter. Two primary methods exist to generate a large-bandwidth free electron laser. One is to use an energy-chirped electron beam \cite{18,19,20,21}, and the other is to utilize space-field correlations in the undulator, such as by injecting a head-to-tail tilted beam into a transverse gradient undulator (TGU) \cite{Prat2016} or a planar undulator \cite{song2018bandwidth}. Using a large energy-chirped electron beam is a simple and natural way to generate broadband XFEL radiation. Recently, the use of a special overcompression \cite{18} scheme to generate a large energy chirp has attracted increasing attention \cite{19,21,24}. In this scheme, electron beams with energy chirps are overcompressed in a bunch compressor, where the heads and tails of beams exchange their positions, and resulting wakefields from the RF structures further increase the energy chirp.

To overcompress electron bunches, the operating parameters of accelerating sections, linearizer, and bunch compressor must be appropriately changed. Optimizing these parameters and determining an affordable XFEL bandwidth are crucial. The optimization goal is to generate electron beams with large energy chirps, high peak currents, good current profiles, and reasonable slice energy spreads. Some of these objectives are contradictory, and thus one must explore global optimal solutions that compromise these objectives. Here, a Pareto-dominance-based multiobjective evolutionary algorithm, NSGA-II \cite{27}, is adopted to maximize the beam energy chirp in an SXFEL, and thus its radiation bandwidth. 

Currently, the SXFEL user facility is under construction at the Shanghai Synchrotron Radiation Facility campus \cite{38}. In the SXFEL user facility, 500 pC electron bunches are generated and accelerated to 130 MeV at the S-band photo-injector and further accelerated by the linac. The main linac consists of three accelerating sections and two bunch compressors. The first S-band accelerating section (L1) is operated at off-crest to induce energy chirp in the bunch, and an X-band linearizer is used to linearize the energy chirp. The first magnetic chicane compresses the electron bunches at an energy of 256 MeV. The subsequent second and third C-band accelerating sections (L2, L3) increase the beam energy to 1.5 GeV. Another chicane lies between L2 and L3 to further compress the bunch. In the large-bandwidth operation mode of the SXFEL, the second magnetic chicane is supposed to be turned off in order to take advantage of strong longitudinal wakefields. The SXFEL has two undulator lines, i.e., a two-stage seeded FEL line to generate fully coherent 3 nm FEL radiation and a SASE line which is proposed to generate large-bandwidth radiation. In the SASE line, 10 in-vacuum undulators with a 16-mm period and 4-m segment length are used. A schematic layout of the facility with a single-stage bunch compressor and SASE line is shown in Fig.\ref{FIG1}. Well-benchmarked codes are used to perform the start-to-end simulations of the SXFEL here. ASTRA \cite{35} is used to track electron bunches in the photo-injector. ELEGANT \cite{Borland2000} is used for simulations of the main linac where collective effects are considered. Generations of the XFEL pulses in the undulator are simulated by GENESIS \cite{Reiche2007Recent}.

\begin{figure}[htp]
\centering 
\includegraphics[width=80mm]{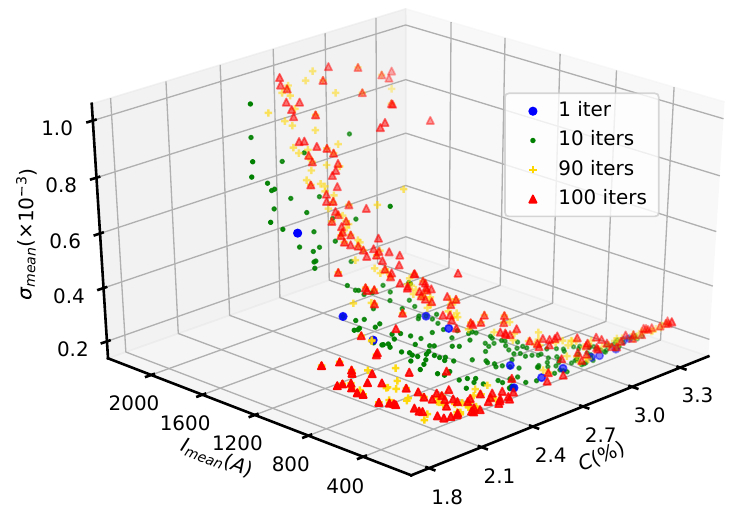} 
\caption{Pareto-optimal fronts at 1, 10, 90, and 100 iterations of the optimization with NSGA-II.} 
\label{fig:Fig2} 
\end{figure}

During optimization, the voltages and phases of the first S-band accelerating section, the X-band linearizer, and the angle of the first bunch compressor are chosen as optimization variables. Constraints of these variables are set according to the corresponding hardware limits of the SXFEL, such as the voltage ranges of the RF structures. Optimization objectives include the current, slice energy spread, and energy chirp, all of which are calculated by the parallel ELEGANT code with 100k macroparticles. An electron beam is divided into 40 slices for calculating these objectives. The current objective ($I_{mean}$) is defined as the average current of the twenty central slices. The slice energy spread objective ($\sigma_{mean}$) represents the average slice energy spread of these 40 slices. The beam energy chirp ($C$) is calculated by the absolute energy difference between the first and last slice. In order to avoid electron beams with poor current profiles and to decrease search areas, additional restrictions have been given during the optimization. For example, if the bunch charge in the central 20 slices is less than 70\% of the total bunch charge, i.e., a poor current profile solution is given the worst fitness values. 

\begin{figure*}[htp]
\includegraphics[width=140mm]{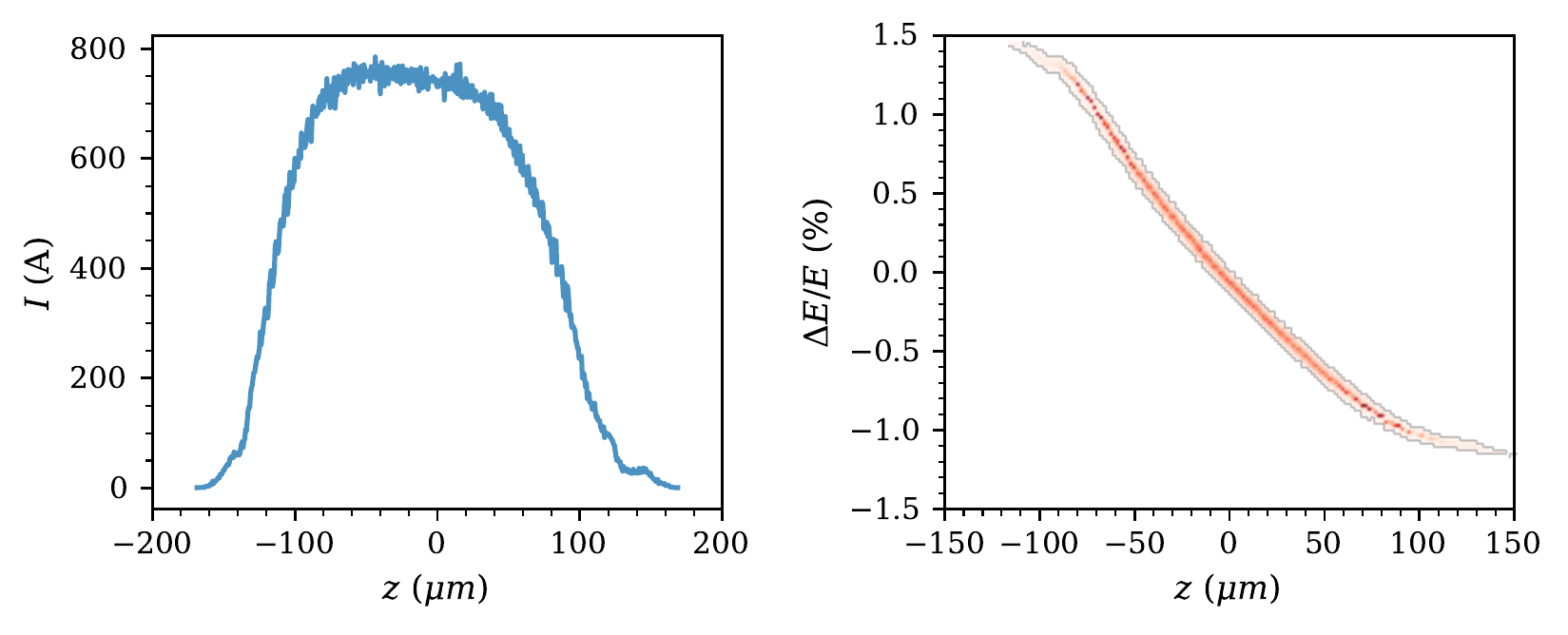}
\caption{Current profile (left) and longitudinal phase space (right) of the chosen electron beam at the entrance of the undulator.} 
\label{FIG3} 
\end{figure*}

\begin{figure*}[htp]
\includegraphics[width=140mm]{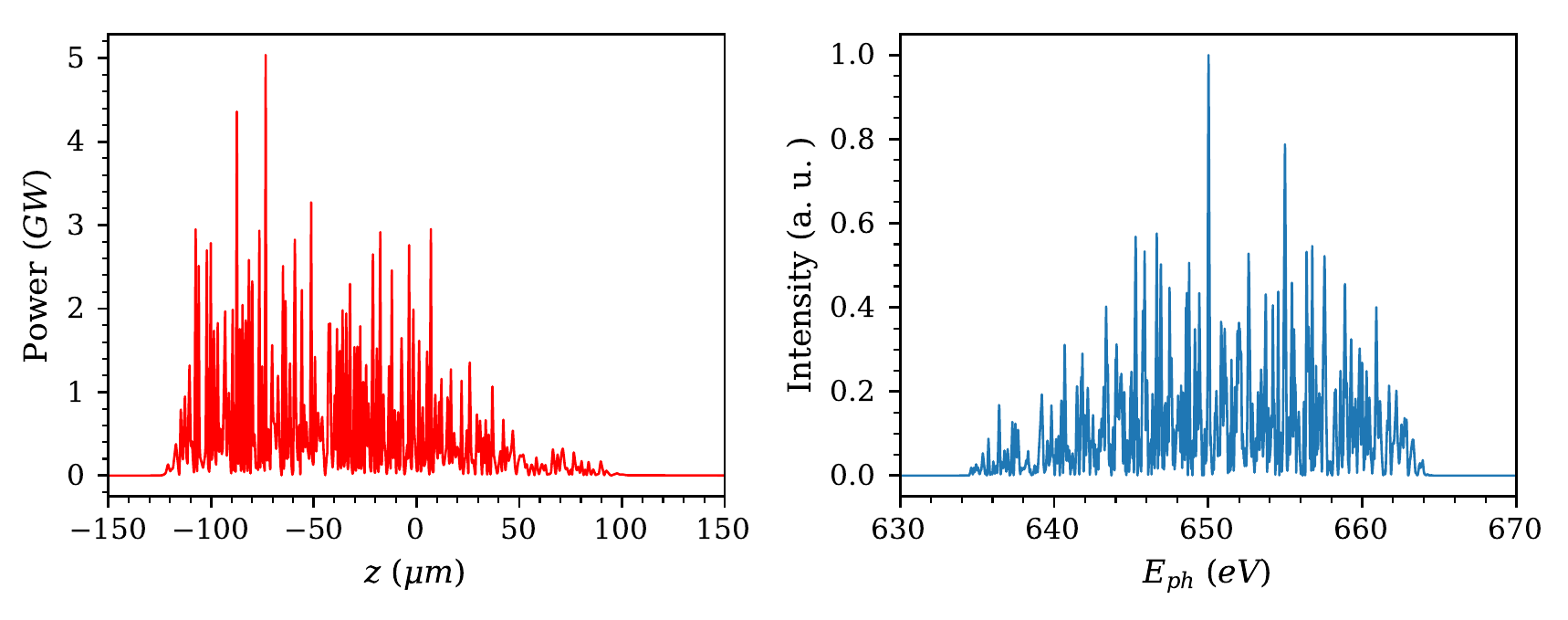} 
\caption{XFEL radiation profile (left) and spectrum (right) at the end of the undulator line for the chosen solution.} 
\label{fig:Fig4}
\end{figure*}

The population size and number of iterations in the algorithm are set to 200 and 100, respectively. Fig.\ref{fig:Fig2} depicts the Pareto-optimal fronts in 1, 10, 90, and 100 generations. The final Pareto-optimal front shows that when the current is between 700 and 1000 A, which is near the normal operation value in the SXFEL, the maximum energy chirp is 2.7\%. Solutions with more than 3\% energy chirp can be achieved when the current is less than 500 A or larger than 2000 A, but these are not suitable for FEL lasing. An acceptable solution from the Pareto front was chosen as an example for the following broadband XFEL-based XAS and XMCD experiment scheme. The beam energy chirp of the chosen solution is 2.51\%, as shown in Fig.\ref{FIG3}, and a quasi-uniform current profile with a 700 A peak current was ensured. The undulator gap was adjusted according to the required central wavelength. Fig. \ref{fig:Fig4} plots the corresponding XFEL radiation profile and spectrum of an FEL pulse at the end of the undulator line. The final FEL pulse energy was 323 $\mu J$, and the full bandwidth including a 2\% cut was 4.4\%, which is near the theoretical value of twice the energy chirp.

\section{Principles and methods of highly efficient XMCD}
\label{sec:proceduces}

As mentioned above, a transmission-based method was chosen for XMCD detection. Combining the energy spectra of the incident and transmitted XFELs, the absorption cross-section can be obtained according to Eq. \ref{eq:absorptionXS}.
\begin{equation}\label{eq:absorptionXS} \sigma_{ab}(E)=-\ln\left(\frac{\Phi_{out}(E)/\eta_{det}(E)}{\Phi_{in}(E)/\eta_{det}(E)}\right)\cdot\frac{1}{N_vl}
\end{equation}
where $\Phi_{out}(E)$ ($\Phi_{in}(E)$) represents the $ E_{ph} $ spectrum of outgoing (incident) photons; $ \eta_{det}(E) $ is the detection efficiency for the XFEL with a given energy E; $ N_v $ is the number density of an atom in the sample material, and $ l $ is the thickness of the target in the incident direction of the XFEL. The XMCD experiment requires simultaneous circular polarization and energy spectrum measurement, which can be achieved by using an e-TOF experiment \cite{Zhang:2017ncj} based on photoelectrons' angular distribution \cite{TRZHASKOVSKAYA2001,Manson1982} and drift time, respectively \cite{TRZHASKOVSKAYA2001}. To measure the $ E_{ph} $ spectrum, the relationship between $ E_{ph} $ and the photoelectron energies $ E_e $ is 
\begin{equation}\label{eq:Eph_TOF}
E_{ph}=E_{e}+E_{b}=\frac{1}{2}m_e(\frac{L}{\Delta t})^2+E_b
\end{equation}
where $ m_e $ is electron's rest mass and $ E_b $ represents the binding energy of the shell originating the photoelectrons \cite{TRZHASKOVSKAYA2001}. Thus, for an e-TOF instrument with a given drift length ($ L $), the outgoing (or incident) $ E_{ph} $ spectrum can be derived from the photoelectrons' times of flight ($ \Delta t $) according to Eq. \ref{eq:Eph_TOF}.

Under strong magnetic fields, electrons excited by fully circularly polarized photons have different transition probabilities due to their different spins, and thus a sample's absorption cross-section varies with the direction of circular polarization ($\mu=\pm$), the so-called XMCD phenomenon. Combining the cross-sections for the XFEL with two opposite circular polarization directions, the XMCD (defined as $\sigma_{\mu_{+}}-\sigma_{\mu_{-}}$) and XAS (defined as $\frac{1}{2}(\sigma_{\mu_{+}}+\sigma_{\mu_{-}})$) can be obtained \cite{Thole1992,Nesvizhskii2000}. According to the famous Sum rule \cite{Thole1992}, the orbital momentum $L_z$ and spin momentum $S_z$ are \cite{Thole1992,Stohr1999}
\begin{align}  
& L_z\propto(\Delta L_2+\Delta L_3) \label{eq:orbit_mag} \\
& S_z+\frac{7}{2}T_z\propto \Delta L_3 - 2\Delta L_2 \label{eq:spin_mag}
\end{align} 
where $ \Delta L_{2} $ ($ \Delta L_{3} $) represents the integration value of the edge $ L_2 $ ($L_3$) in the XMCD spectrum, and $ T_z $ is the magnetic dipole term related to anisotropic charge, which can be neglected with high crystal symmetries and small spin-orbit splitting \cite{Stohr1995,Nesvizhskii2000}. The ratio of $ L_z $ to $ S_z $ (branching ratio, $ BR $) is usually used to infer electronic structure \cite{Nesvizhskii2000,Laguna-Marco2010,Laan1990} by Eq. \ref{eq:rls}.
\begin{equation}\label{eq:rls} 
BR=L_z/S_z=\frac{\Delta L_2+\Delta L_3}{\Delta L_3 - 2\Delta L_2} 
\end{equation} 
In order to evaluate the performance of the XMCD experiment, the relative deviation of $ BR $ will be used for further analysis. As in a recent time-resolved XMCD experiment, a pump laser was used to excite target atoms and a fully circularly polarized XFEL pulse with a delay was used for measuring the absorption cross-section when atoms de-excited, from which, critical microscopic properties such as electronic structure could be inferred \cite{Higley2016}. By changing the pump-probe time interval, atomic de-excitation can be recorded instantaneously \cite{Higley2016}.

\begin{figure*}[htp] 
\centering 
\includegraphics[width=\linewidth]{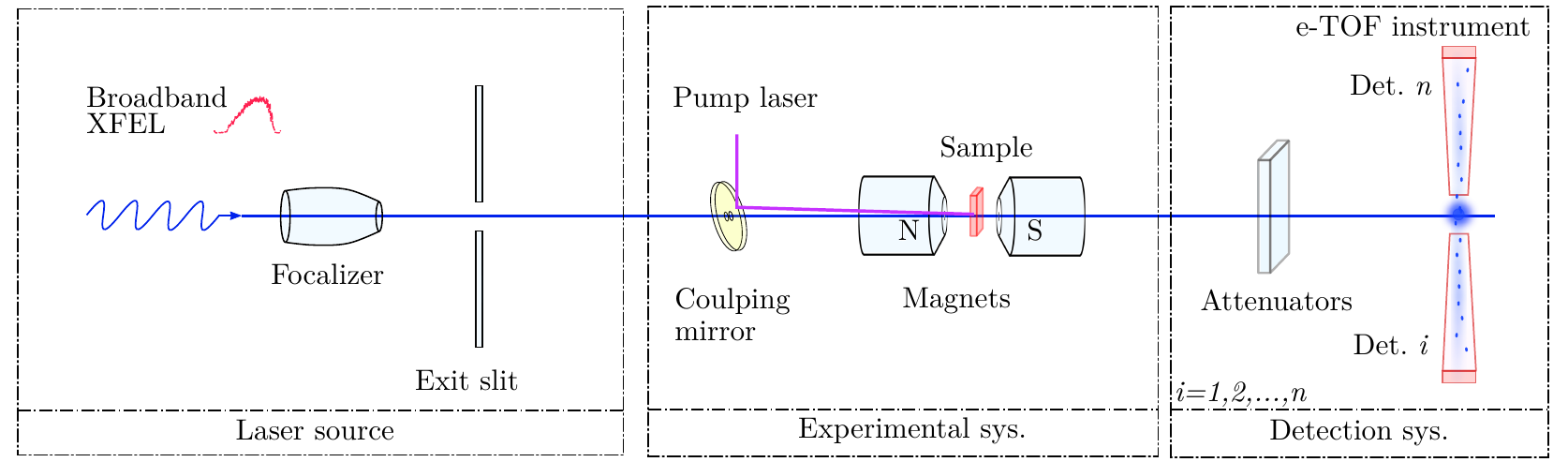}
\caption{Block diagram of the experiment} 
\label{fig:arrangement} 
\end{figure*}

The proposed experiment facilities are designed to contain four parts: the laser source, experimental system, detection system, and electronics system. A block diagram of the experimental arrangement is shown in Fig. \ref{fig:arrangement}. The laser source provides broadband XFEL pulses which fluctuate from shot to shot. However, a steady energy spectrum can be obtained by superposing an $ E_{ph} $ spectrum with a proper number of pulses. Thus, two $ E_{ph} $ superposed spectra with sufficient pulses close to each other can be used to measure the absorption spectra. One is used for measuring incident spectra and the other can be used for measuring transmitted spectra after light passes through the sample. Then, the absorption cross-section for the XFEL with a given circular polarization can be derived according to Eqs. \ref{eq:absorptionXS}, \ref{eq:orbit_mag}, \ref{eq:spin_mag} and \ref{eq:rls}. The experimental system is where the XFEL interacts with the target and is used to measure the $ E_{ph} $ spectrum by using the e-TOF instrument. After a short delay caused by the pump laser, which transfers target atoms to an excited state, a circularly polarized XFEL pulse interacts with the sample material. By changing the pump-probe delay and using enough pulses, the XMCD during the entire de-excitation process can be monitored. According to Eq. \ref{eq:absorptionXS}, spin and orbital magnetism, which are related to the electron state densities and spin directions at different de-excitation times, can be obtained. The electronics system contains a time-to-digital converter (TDC) and a waveform digitizer for pulse-shape sampling. Signal processing is described briefly below. Once photoelectrons reach the micro-channel plate electron detector, a signal is generated and sent to the discriminator. The TDC is triggered by a start signal from the XFEL source and stopped by a signal from the discriminator, recording the flight time of the earliest photo-electrons. Meanwhile, a waveform digitizer with a sampling interval is triggered to record the signal after a proper delay, aiming to cover the whole spectrum. The sampled signal from the waveform digitizer is then analyzed online and de-composed to obtain the $ E_{ph} $ spectrum by deconvolution.

Generally, noise from phenomena such as XFEL beam fluctuation and photoelectron drift lead to lower quality deconvolutions. Compared to our previous research \cite{Zhang:2017ncj}, an improved method using Wiener deconvolution \cite{Wiener1949} has been used to retrieve $ E_{ph} $ from a signal pulse robustly, reducing the influence of the noise significantly. In the following, this method will be described. First, a digitized signal is adjusted by various time offsets, e.g., due to electronics cables and detector response delays. Second, the adjusted digitized signal is interpolated with a smaller time interval for signal deconvolution, because the sampling period is much greater than the real interval between two arrival photoelectrons. Finally, the Wiener algorithm is used to reconstruct the photoelectron spectrum from the signal pulse shape as described below.

The interpolated time-of-flight (TOF) signal shape $ P(t) $ can be described by Eq. \ref{eq:Pt_timedomain} in the time domain:
\begin{equation}
\label{eq:Pt_timedomain}
P(t)=S(t)*[N(t)*N_s(t)]+e(t)
\end{equation}
where $ * $ represents the convolution operation; $S(t)$ is the pulse shape of an MCP detector for a single electron without delay; $N(t)$ is the theoretical flight time of a photoelectron from a photon with a given energy in the vacuum; $ N_s(t) $ is the distribution of photon electrons in the time domain (without an offset of the earliest flight time); $ F(t) $ is the signal pulse shape for an impulse in time domain, which can be described in the shape of a Log-Gaussian distribution \cite{Zhang:2017ncj}; and $ e(t) $ is an error term due to noise and statistical fluctuations. In the frequency domain, Eq. \ref{eq:Pt_timedomain} can be written
\begin{equation}
\label{eq:Pt_freqdomain}
P(f)=F(f)N (f) + e(f).
\end{equation}
According to the Wiener deconvolution algorithm, the estimated distribution of the theoretical TOF in vacuum ($ \hat{N}(t) $) can be obtained by applying a filter $ W(f) $ to $ P(f) $, which satisfies the relation as shown in Eq.\ref{eq:Wf}.
\begin{equation}\label{eq:Wf}
\hat{N}(t)=\mathcal{F}^{-1}[W(f)\cdot P(f)]
\end{equation}
where $ \mathcal{F}^{-1} $ represents the inverse Fourier transform. Analytically, an optimal $ W(f) $ can be written \cite{Wiener1949}
\begin{equation}\label{eq:optimalWf}
W(f)=\frac{F^{*}(f)}{|F(f)|^2+|e(f)|^2/|N(f)|^2}
\end{equation}
From Eq.\ref{eq:optimalWf}, the impulse response of a detection instrument (e-TOF) and the power spectrum of the signal-to-noise ratio should be known. The former can be measured using a conventional XFEL and monochromators, while the latter can be estimated by the local variance of the signal with a proper window, or by a known spectrum measured by another conventional spectrometer under similar conditions. Finally, the $ E_{ph} $ spectrum can be obtained from the estimated flight time $\hat{N}(t)$ according to Eq. \ref{eq:Eph_TOF}. 

Here, the detail methodology is described. According to the experimental method described above, procedures include two stages: preparation and operation. The preparation stage aims to calibrate the e-TOF instrument. The XFEL source in its conventional mode (narrow bandwidth) and the monochromator were used to measure the average impulse response of the e-TOF instrument $ F(t) $ and the detection efficiency $ \eta_{det}(E_{ph}) $, by choosing the proper number of energy points in the energy range and B-spline interpolation. The preparation stage is only performed once and is not required when samples are changed. The operation stage aims to measure the absorption cross-sections of circularly polarized photons in two polarization directions, and consists of the following steps. First, operational tuning of the broadband mode must be performed to modulate the XFEL to provide fully circularly polarized photons covering the entire energy range, both of which can be diagnosed by the e-TOF instrument simultaneously \cite{Zhang:2017ncj}. Second, the target must be placed into the experimental container with a proper magnetic field. By using sufficient XFEL pulses and setting a proper pump-probe delay, the electrons' time-of-flight signal shapes for the XFEL can then be recorded, which are then analyzed online and used to obtain the absorption spectrum $E_{ph}$. Third, the time-resolved absorption spectrum for XFEL must be measured when fully circularized in a particular direction by changing the pump-probe interval and repeating step 2. Then, the TOF signal must be recorded using the same number of XFEL pulses after removing the target, and the incident $ E_{ph} $ spectrum can be reconstructed accordingly. Combining with absorption spectrum in step 3, the corresponding time-resolved absorption cross-section can be obtained. Then, the time-resolved cross-section for another circular polarization direction is measured by reversing the circular polarization direction or, equivalently, the direction of magnetic field \cite{Stohr1999}. Finally, the XMCD and XAS spectra can be calculated according to the measured absorption spectra in the two circularly polarized directions. Furthermore, orbital and spin-magnetic properties can be calculated with Eq. \ref{eq:orbit_mag} and Eq. \ref{eq:spin_mag}.

\section{Numerical modeling and sensitivity analysis} \label{sec4} 
\begin{figure}[htp]
\centering
\includegraphics[width=80mm]{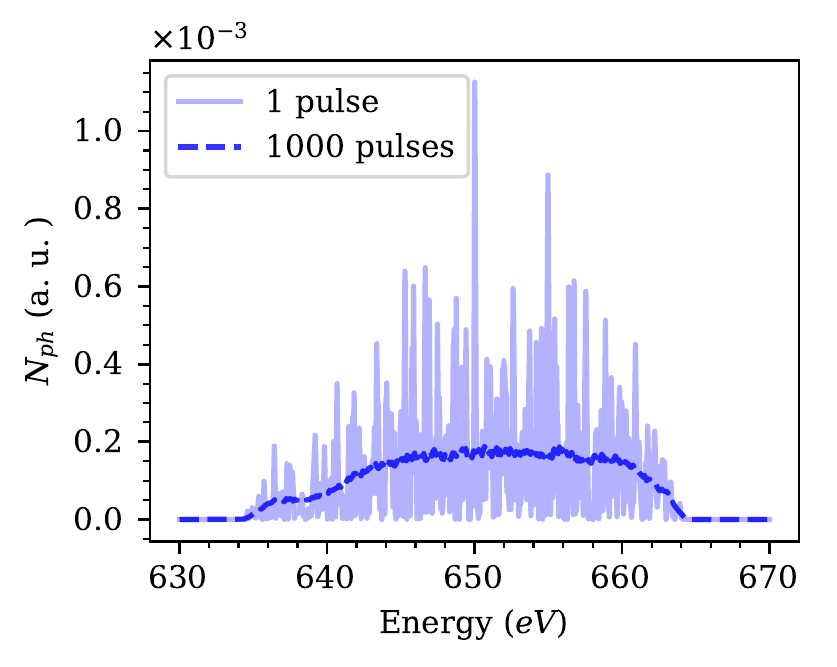}
\caption{Superposed $ E_{ph} $ spectrum with different number of pulses.}
\label{fig:Energy-N} 
\end{figure}

\begin{figure}[htp]
\includegraphics[width=80mm]{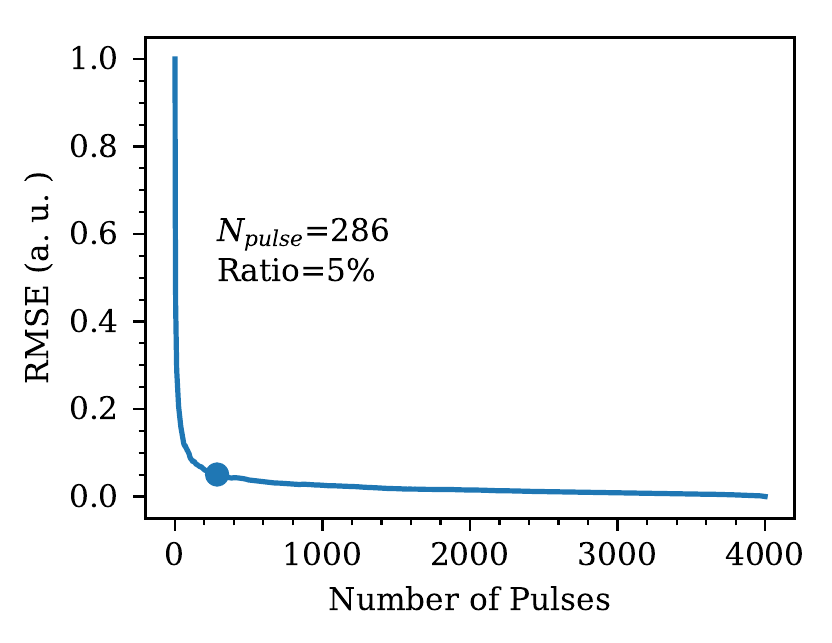} 
\caption{Relationship between the RMSE and number of XFEL pulses. The $ E_{ph} $ spectrum obtained with 4000 pulses is regarded as a steady spectrum.}
\label{fig:saturation-eff-5percent}
\end{figure}

In order to clearly describe the proposed XAS/XMCD method driven by a broadband XFEL pulse, an example with an MnO sample is numerically illustrated based on the broad-bandwidth XFEL discussed in section \ref{sec2}. We chose a solution from the optimization above to generate large energy-chirped electron bunches. The current profile and longitudinal phase space of the electron beam are shown in Fig. \ref{FIG3}, and the power and energy spectra of a single broadband XFEL pulse are shown in Fig. \ref{fig:Fig4}. As mentioned above, considering the shot-to-shot fluctuation of the SASE pulse spectrum, a steady FEL spectrum can be obtained by superposing multiple SASE pulses, as shown in Fig.\ref{fig:Energy-N}. The root-mean-square-error (RMSE) was used to evaluate the stability and accuracy of the superposed FEL spectrum, defined as
\begin{equation}
\label{RMSE}
RMSE=\left[(E_{ph,st}-E_{ph})^2/n\right]^{1/2}
\end{equation}
where $E_{ph,st}$ represents the steady spectrum and $ n $ is the number of bins. For convenience, the superposed $ E_{ph} $ spectrum of 4000 pulses is regarded as a steady spectrum. The relationship between the RMSE and number of pulses ($N_{ph}$) is shown in Fig. \ref{fig:saturation-eff-5percent}, from which the RMSE of the spectrum superposed by 286 pulses was decreased to $ 5\% $ compared with $ E_{ph} $ spectrum of a pulse, suggesting that a potentially practical number of pulses is about 1000 pulses. Thus, the number of pulses was set to 286 temporarily for this baseline calculation.

\begin{figure*}[tph] 
\centering 
\includegraphics[width=\linewidth]{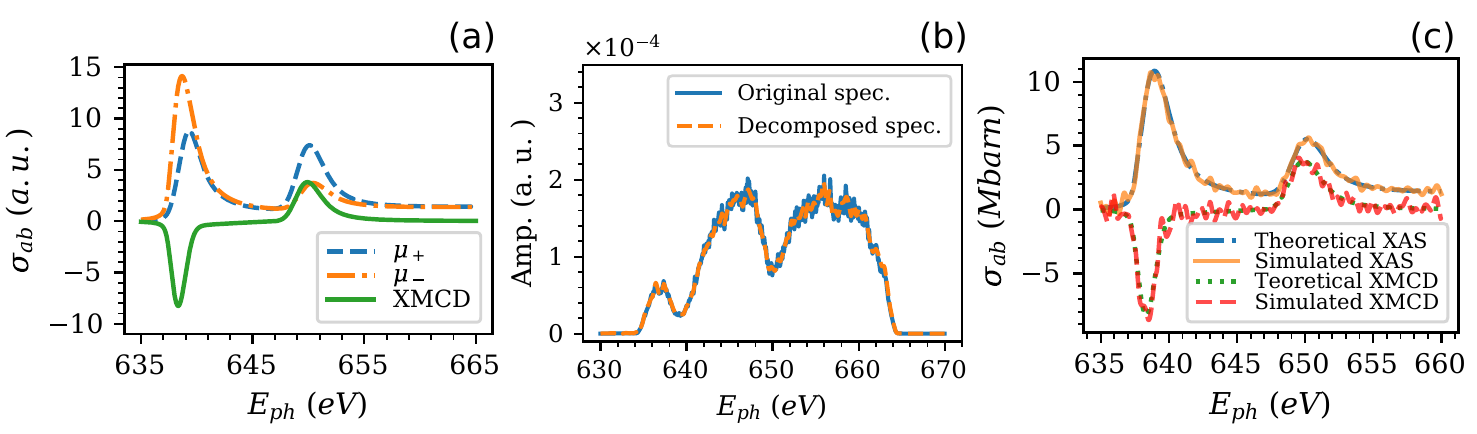}
\caption{Simulation results using our numerical model.
a) Simulated absorption cross-section of MnO.
b) Comparison of decomposed transmission $ E_{ph} $ and original simulated spectra.
c) Comparison of decomposed XMCD/XAS spectra with theoretical XMCD/XAS in Fig \ref{fig:SimulationResult}(a). } \label{fig:SimulationResult}
\end{figure*}

The calculation of the typical cross-section of MnO in the ground state for both left- and right-circularly polarized FELs was performed in the framework of FDMNES \cite{Joly2009Self}. The MnO was in the ground state and the crystal space group was Fm3m ($a=b=c=4.436$ \AA and$ \alpha=\beta=\gamma=90^\circ $). The magnetic field was in the $z$ direction, perpendicular to the plane of the X-ray's circular polarization. Gaussian broadening ($ \mu=\ 0.1 eV $, $ \sigma=0.05\ eV $) and a uniform broadening of $0.8 eV$ due to complex energies were implemented to be consistent with the experimental absorption cross-section from the Shanghai Synchrotron Radiation Facility. Other influences such as self-absorption and birefringence were also taken into consideration. A typical dichroism cross-section calculation with the MnO sample is shown in Fig. \ref{fig:SimulationResult}(a), and is used to calculate absorption spectrum for the demonstration of the experimental method.

As mentioned above, polarization diagnosis and energy spectrum measurement can be performed by using an e-TOF instrument, whose numerical model (in the framework of Geant4 \cite{Chauvie2001Geant4}) has been designed thoroughly in our previous research \cite{Zhang:2017ncj}. In this work, the numerical model and optimal design were reused for modeling and analysis. The enhancement of the e-TOF signal decomposition method described in Eq. \ref{eq:Pt_timedomain} to Eq. \ref{eq:optimalWf} for $ E_{ph} $ measurement was implemented in this work. According to Eq. \ref{eq:absorptionXS} and the description of the preparation stage, a measurement of the detection efficiency curve for different energies is required. In simulation, the detection efficiency was measured by using several XFEL pulses with high coherencies selected by the monochromator in conventional mode, whose energy was selected by using a Gaussian distribution with mean of 0.026 eV and a standard deviation that varied in energy.

Other critical parameters in the simulation are listed in Table \ref{tab:parameters}. To evaluate experimental performance, the RMSE of XMCD (or XAS) and the relative deviation of the branching ratio ($ BR $) were used to evaluate statistical error and the measurement precision, respectively. Because of an MnO crystal's high symmetry and small spin-orbit splitting, the magnetic dipole term $ T_z $ can be neglected in $BR$ calculation \cite{Stohr1995,Nesvizhskii2000}. The result of the simulation is shown in Fig.\ref{fig:SimulationResult}. The relative deviation of $ BR $ is 0.98\% and the RMSE of the XMCD and XAS spectra reached 0.62 and 0.42 respectively, essentially satisfying the experimental requirements. Owing to the limited number of detected photoelectrons (about 9000 per detector per pulse) and low XFEL pulse repeating frequency (50 Hz), the influence of vacuum space-charge effects in detection can be ignored.

\begin{table}[htpb]
\centering
\caption{Critical parameters of the numerical model}
\begin{ruledtabular}
\begin{tabular}{p{1.2cm}ll}
Params. & Value                          & Descriptions                                           \\
\colrule

L                 & 400 (mm)                       & The drift length of photo-electrons.                   \\
$\Delta T_s$      & 50 (ps)                        & The Sampling time interval.                                 \\
$N_{p}$           & 286                            & The number of XFEL pulses.                              \\
$d_m$             & $1.58\times10^{-7}$ ($g/cm^2$) & The mass thickness of the sample.                      \\
Sample            & MnO                            & The Sample material.                            \\
D                 & 42 (mm)                        & The effective diameter of electron detectors.           \\
Gas               & $O_2$                          & Target gas in the e-TOF instrument. \\
$ E_b $           & 539.43 eV                      & The binding energy of target gas.      
\end{tabular} \label{tab:parameters}

\end{ruledtabular}
\end{table}

Multiple critical parameters can affect XMCD/XAS measurements, which prefers a steady superposed spectrum for incident and transmitted $ E_{ph} $ measurement and a high-precision pulse shape digitization. A steady superposed spectrum can be obtained by using sufficient XFEL pulses as mentioned previously. In addition, a high-precision pulse-shape digitization can be achieved with a small digitization interval and a long pulse time duration, which suggests the use of a long drift length. In the following, the influence of three critical parameters (the number of pulses, digitization interval, and photoelectrons' drift lengths) on the experimental performance will be analyzed by changing each parameter value individually while holding others fixed at values listed in Table \ref{tab:parameters}.

\begin{figure*}[htp] 
\centering 
\includegraphics[width=\linewidth]{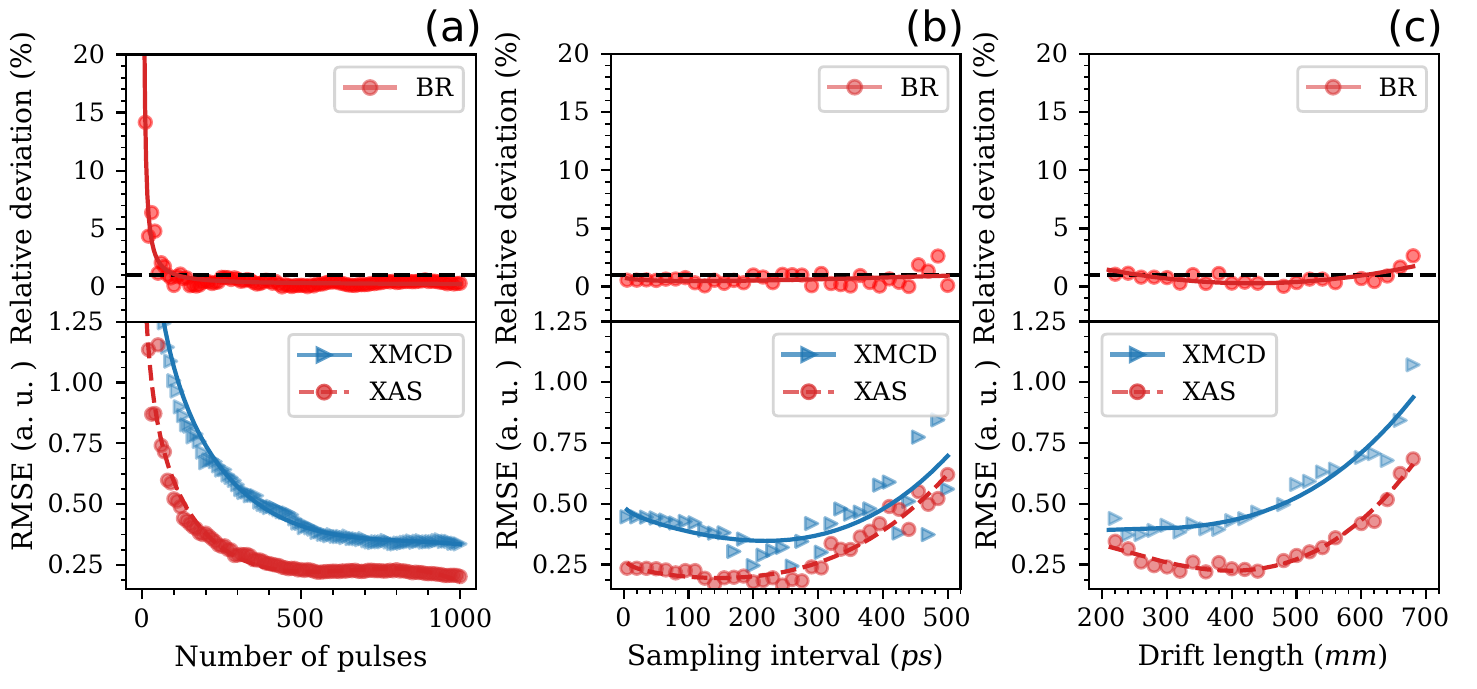}
\caption{Sensitivity analysis of experimental performance with (a) different number of pulses; (b) different sampling interval; (c) different drift length.}
\label{fig:sensitive_analysis}
\end{figure*}

For a given number of pulses, large quantities of XFEL pulses can reduce the fluctuations of the electric field in undulators and statistical error due to the number of detected electrons. In order to determine the minimal XFEL shots, the relationship between the number of pulses and optimization objects were analyzed, with the result displayed in Fig. \ref{fig:sensitive_analysis}(a). The minimal number of pulses required for achieving the design goal (relative deviation of branch ratio $\sim 1\%$) is 200. However, to balance the RMSE value, 400 pulses were chosen as the optimized value.

For the sampling interval, a small digitization period is preferable for a precise e-TOF signal. However, smaller sampling intervals lead to a higher noise sensitivity, while larger sampling intervals produce inaccurate digitization. The related analysis is demonstrated in Fig. \ref{fig:sensitive_analysis}(b). From the simulation, the optimal sampling period is around 250 ps, where the RMSE is the smallest, while the relative deviation of the measured branch ratio changed little due to the smoothness of integration.

A long drift length is preferable, since it converts small energy intervals into measurable time intervals. However, a larger drift distance also induces larger statistical error and is more strongly influenced by the Earth magnetic field. The influence of different drift lengths is shown in Fig. \ref{fig:sensitive_analysis}(c), wherein an optimal drift length of $ \sim 450 {\ \rm mm}$ with minimal RMSE is achieved while the branch ratio changes only a little.

By implementing the optimal design outlined above, a relative deviation of the BR was reduced to $0.79\%$ and the RMSE values of 0.38 and 0.27 were achieved for the XMCD and XAS spectra, respectively. Since only 400 consecutive shots were needed, the experiment can be completed within tens of minutes with one or two machine adjustments.

\section{Single-shot XAS/XMCD}
\label{sec5}
\begin{figure*}[t] \centering
\includegraphics[width=\linewidth]{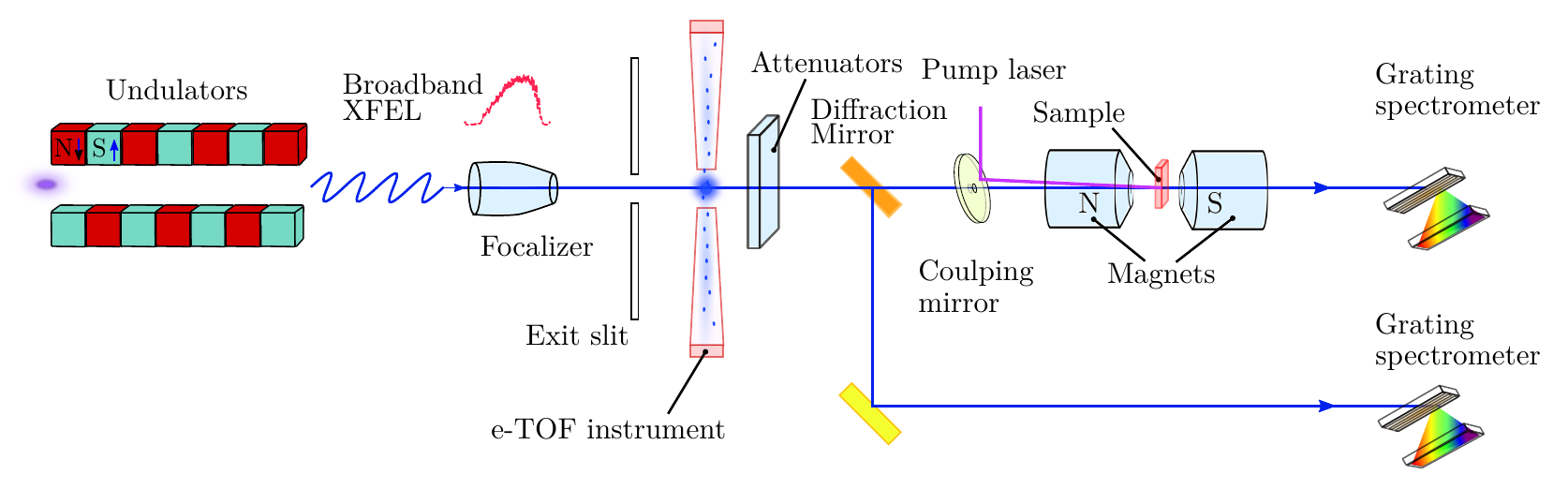}
\caption{Apparatus blocks for the single-shot time-resolved absorption cross-section measurement.}
\label{fig:arrangement-beam-diffraction}
\end{figure*}

Single-shot XAS/XMCD is extremely appealing for fast measurements when sample materials are too sensitive to allow numerous pulses. For a single pulse, the e-TOF spectrometer utilizes only a small fraction of photons. The electron yield is sufficient for $ E_{ph} $ diagnosis using a single shot but not for precise single-shot $ E_{ph} $ measurement. Grating spectrometers are capable of measuring broadband spectrum of each pulse, due to the highest utilization efficiency of photons compared to an e-TOF spectrometer. By using gratings and beam diffraction techniques, it is possible to perform pulse-to-pulse time-resolved XAS measurement. Furthermore, the XMCD spectra can be obtained with at least two XFEL pulses and a reverse of the direction of the magnetic field.

The apparatus arrangement is shown as Fig. \ref{fig:arrangement-beam-diffraction}. Here, the e-TOF instrument is only used for monitoring circular polarization and the $ E_{ph} $ spectrum, while the grating spectrometers are used for precise $ E_{ph} $ measurement with the XFEL. For a single shot, a circularly-polarized broadband XFEL pulse with a pump-probe time delay was diffracted into two beams. In one, the XFEL interacts with the target sample and the transmitted absorption spectrum is calibrated precisely with the grating spectrometer. In the other, the XFEL is used for measuring the incident XFEL energy spectra. Combining these two spectra, the absorption cross-section in one polarization direction can be obtained. Then, the direction of the magnetic field is reversed and the absorption cross-section for another polarization direction can be measured along with the XMCD/XAS spectra.

In this way, a single shot with a proper pump-probe delay can be used for absorption spectrum measurement, and two pulses are required for XMCD and XAS spectra. A more complicated apparatus and optical path can be designed for future single-shot time-resolved XMCD/XAS spectra measurements, and requires simultaneous calibration of the absorption cross-section for the XFEL in two circular polarization directions. A possible design might inject an XFEL pulse from each of two sides of a sample in the same magnetic field and measure the transmission $ E_{ph} $ spectrum for each. Another possible design might use two pulses of multicolored emitted XFEL light with a delay and a fast magnetic field reverse.

\section{Conclusions} \label{sec6}
The feasibility of a highly-efficient method for an XMCD experiment driven by broadband X-ray free-electron lasers has been proposed and numerically validated. To determine a satisfactory radiation bandwidth region and corresponding optimal operation parameters in the SXFEL user facility, an evolutionary multiobjective optimization algorithm was utilized to maximize the XFEL bandwidth. With hundreds of broadband FEL pulses ($\approx$ 30 eV) covering an entire absorption spectrum, an e-TOF-based spectrometer, and a robust e-TOF signal post-decomposition, the XAS and XMCD can be measured with the same FEL machine in several minutes without resorting to tedious FEL wavelength tuning and other time-consuming machine operations.

The proposed method can be further improved to measure single-shot absorption spectra with X-ray diffraction splitting, dramatically reducing the risk of damage for highly radioactive-sensitive sample materials.

\begin{acknowledgments} 
The authors would like to thank Linjuan Zhang and Shuo Zhang for helpful discussions on X-ray spectroscopy. This work was partially supported by the National Natural Science Foundation of China (11775293), the National Key Research and Development Program of China (2016YFA0401900), the Fundamental Research Funds for the Central Universities (Grant No. xjj2017109), the Young Elite Scientist Sponsorship Program by CAST (2015QNRC001), and the Ten Thousand Talent Program.
\end{acknowledgments}

\bibliography{Broadband_FEL_Latex_APS} 
\end{document}